\documentclass[camera]{jpaper}

\usepackage{multirow}
\usepackage{mathtools}
\usepackage{xspace}
\usepackage{textcomp}
\usepackage{algorithm}
\usepackage{algorithmic}
\usepackage{subfig}
\usepackage{xfrac}
\usepackage{indentfirst}
\usepackage{subscript}
\usepackage[nocompress]{cite}
\usepackage{balance}
\usepackage{array}
\usepackage{gensymb}

\makeatletter
\let\MYcaption\@makecaption
\makeatother

\makeatletter
\let\@makecaption\MYcaption
\makeatother

\usepackage{fixltx2e}
\usepackage{dblfloatfix}
\usepackage[nolessnomore, italic]{mathastext}
\usepackage[T1]{fontenc}
\usepackage[usenames,dvipsnames,svgnames,table]{xcolor}
\usepackage[normalem]{ulem}
\usepackage{enumitem}
\usepackage{setspace}
\usepackage{footmisc}
\usepackage{fancyhdr}
\usepackage{authblk}
\usepackage[us,12hr]{datetime}
\usepackage[keeplastbox]{flushend}

\newif\ifcameraready
\camerareadytrue

\newcommand{\versionnum}[0]{5}

\ifcameraready
\else
\fi

\ifcameraready
  \newcommand{\todo}[1][]{}
  \newcommand{\ch}[0]{}
\else
  \newcommand{\todo}[1][]{\textbf{\fcolorbox{black}{red}{\color{white}{TODO}}} \underline{$\overline{\hbox{\emph{#1}}}$}}
  \newcommand{\ch}[1]{{\color{BrickRed} #1}}
\fi

\usepackage{tikz}

\newcounter{hours}
\newcounter{minutes}

\newboolean{timeofmake}
\setboolean{timeofmake}{false}

\newcommand{\ignore}[1]{}

\newcommand{\new}[0]{}

\newcommand{\intersub}[0]{{Subarray Access Refresh Parallelization}\xspace}

\newcommand{\is}[0]{{\text{SARP}}\xspace}
\newcommand{\sarp}[0]{{\text{SARP}}\xspace}

\newcommand{\warplong}[0]{{\mbox{write-refresh} parallelization}\xspace}
\newcommand{\Warplong}[0]{{\mbox{Write-refresh} parallelization}\xspace}
\newcommand{\darp}[0]{{DARP}\xspace}
\newcommand{\darplong}[0]{{Dynamic Access Refresh Parallelization}\xspace}

\newcommand{\ooolong}[0]{{\mbox{out-of-order} per-bank refresh}\xspace}

\newcommand{\ib}[0]{{\darp}\xspace}

\newcommand{\combo}[0]{{DSARP}\xspace}

\newcommand{\refab}[0]{{\small\emph{$REF_{ab}$}}\xspace}
\newcommand{\refpb}[0]{{\small\emph{$REF_{pb}$}}\xspace}
\newcommand{\trefi}[0]{{\small\emph{$tREFI_{ab}$}}\xspace}

\newcommand{\trfc}[0]{{\small\emph{$tRFC_{ab}$}}\xspace}

\newcommand{\fgr}[0]{{\small\emph{FGR}}\xspace}

\newcommand{\caprefab}[0]{\small{\textbf{\textit{REF\textsubscript{ab}}}}\xspace}
\newcommand{\caprefpb}[0]{\small{\textbf{\textit{REF\textsubscript{pb}}}}\xspace}

\newcommand{\figputHW}[3]{
\begin{figure}[h]
\begin{minipage}{\linewidth}
\footnotesize 
\begin{center}
\includegraphics[width=1.0\linewidth]{plots/#1}
\end{center}
\vspace{-0.1in}
\caption{#2 \label{fig:#1}}
\end{minipage}
\end{figure}
}

\newcommand{\figputHS}[3]{
\begin{figure}[h]
\begin{minipage}{\linewidth}
\begin{center}
\includegraphics[scale=#2]{plots/#1}
\end{center}
\vspace{-0.15in}
\caption{#3 \label{fig:#1}}
\end{minipage}
\end{figure}
}

\newcommand{\figputGHS}[3]{
\begin{figure}[h]
\begin{minipage}{\linewidth}
\begin{center}
\includegraphics[scale=#2]{gnuplots/#1}
\end{center}
\vspace{-0.1in}
\caption{#3 \label{fig:#1}}
\end{minipage}
\end{figure}
}

\newcommand{\figref}[1]{Figure~\ref{fig:#1}}

\begin{document}

\title{Reducing DRAM Refresh Overheads\\with Refresh--Access Parallelism}

\author{%
{Kevin K. Chang$^{1,2}$}%
\qquad%
{Donghyuk Lee$^{3,2}$}%
\qquad%
{Zeshan Chishti$^{4}$}%
\vspace{2pt}\\
{Alaa R. Alameldeen$^{4}$}%
\qquad%
{Chris Wilkerson$^{4}$}%
\qquad%
{Yoongu Kim$^{2}$}%
\qquad%
{Onur Mutlu$^{5,2}$}}%
\affil{%
{\it%
$^1$Facebook\qquad%
$^2$Carnegie Mellon University%
\qquad%
$^3$NVIDIA Research\qquad%
$^4$Intel Labs\qquad%
$^5$ETH Z{\"u}rich}%
}

\maketitle

\begin{abstract}
\ignore{Modern DRAM cells are periodically refreshed to prevent data loss due
to leakage. Commodity DDR (double data rate) DRAM refreshes cells at
the rank level. This degrades performance significantly because it
prevents an entire DRAM rank from serving memory requests while being
refreshed. DRAM designed for mobile platforms, LPDDR (low power DDR)
DRAM, supports an enhanced mode, called per-bank refresh, that
refreshes cells at the bank level. This enables a bank to be accessed
while another in the same rank is being refreshed, alleviating part of
the negative performance impact of refreshes. Unfortunately, there are
two shortcomings of per-bank refresh employed in today's systems.
First, we observe that the per-bank refresh scheduling scheme does not
exploit the full potential of overlapping refreshes with accesses
across banks because it restricts the banks to be refreshed in a
sequential round-robin order. Second, accesses to a bank that is being
refreshed have to wait.}

\new{This article summarizes the idea of ``refresh--access parallelism,'' which was
  published in HPCA 2014~\cite{chang-hpca2014}, and examines the work's significance
  and future potential. The overarching objective of our HPCA 2014 paper is to reduce the
  significant negative performance impact of DRAM refresh with intelligent memory
  controller mechanisms.}

To mitigate the negative performance impact of DRAM refresh, our HPCA 2014 paper
proposes two complementary mechanisms, \ib (\darplong) and \is (\intersub). The
goal is to address the drawbacks of \new{state-of-the-art per-bank refresh
  mechanism} by building more efficient techniques to parallelize refreshes and
accesses within DRAM. First, instead of issuing per-bank refreshes in a
round-robin order, as it is done today, \ib issues per-bank refreshes to idle
banks in an out-of-order manner. Furthermore, \ib proactively schedules
refreshes during intervals when a batch of writes are draining to DRAM. Second,
\is exploits the existence of mostly-independent {\em subarrays} within a bank.
With minor modifications to DRAM organization, it allows a bank to serve memory
accesses to an idle subarray while another subarray is being refreshed. Our
extensive evaluations on a wide variety of workloads and systems show that our
mechanisms improve system performance (and energy efficiency) compared to three
\mbox{state-of-the-art} refresh policies, and their performance benefits increase
as DRAM density increases.

\end{abstract}

\section{Introduction}

Modern main memory is predominantly built using \emph{dynamic random
  access memory} (DRAM) cells. A DRAM cell consists of a capacitor to
store one bit of data as electrical charge. The capacitor leaks charge
over time, causing stored data to change. As a result, DRAM requires
an operation called \emph{refresh} that periodically restores
electrical charge in DRAM cells to maintain data integrity.

There are two major ways refresh operations are performed in modern DRAM
systems: {\em all-bank refresh (or, rank-level refresh)} and {\em per-bank
  refresh}. These methods differ in what levels of the DRAM hierarchy refresh
operations tie up. A modern DRAM system is organized as a hierarchy of ranks
and banks. Each rank is composed of multiple banks. Different ranks and banks
can be accessed independently. Each bank contains a number of rows (e.g., 16-32K
in modern chips). Because successively refreshing {\em all} rows in a DRAM chip
would cause very high delay by tying up the entire DRAM device, modern memory
controllers issue a number of refresh commands that are evenly distributed
throughout the refresh
interval~\cite{jedec-ddr3,jedec-lpddr3,liu-isca2013,liu-isca2012,nair-hpca2013}.
Each refresh command refreshes a small number of rows.\footnote{The time between
  two refresh commands is fixed to an amount that is dependent on the DRAM type
  and temperature.
  \new{We refer the reader to our prior
 works~\cite{chang-sigmetrics2016, kim-isca2012, lee-hpca2013,
      lee-hpca2015, kim-micro2010, kim-hpca2010,chang-hpca2016, hassan-hpca2016,
      chang-sigmetrics2017, lee-sigmetrics2017, lee-taco2016, lee-pact2015,
      liu-isca2012, liu-isca2013, patel-isca2017, chang-hpca2014,
      seshadri-micro2013, seshadri-micro2017, hassan-hpca2017, kim-cal2015,
      kim-isca2014, kim-hpca2018} for a detailed background on
  DRAM.}} The two common refresh methods of today differ in where in
the DRAM hierarchy the rows refreshed by a refresh command reside.

In {\em all-bank refresh (\refab)}, employed by both commodity DDR and LPDDR
DRAM chips, a refresh command operates at the rank level: it refreshes a number
of rows in {\em all} banks of a rank concurrently. This causes every bank within
a rank to be unavailable to serve memory requests until the refresh command is
complete. Therefore, it degrades performance significantly\ch{~\cite{liu-isca2012,
    mukundan-isca2013, nair-hpca2013, stuecheli-micro2010, chang-hpca2014,
    baek-tc2014, patel-isca2017}}.

\begin{figure}[h]
\begin{center}
\includegraphics[width=\linewidth]{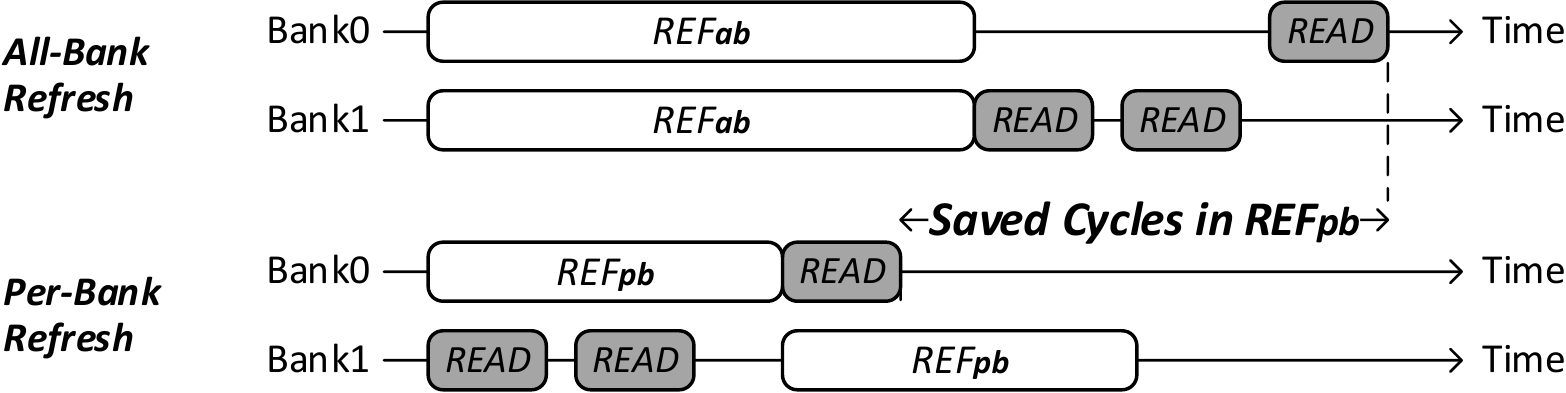}
\end{center}
\caption{Service timelines of all-bank and per-bank refresh. \new{Adapted
    from~\cite{chang-hpca2014}.}}
\label{fig:per-bank-refresh-timeline}
\end{figure}

An alternative method is to perform refresh operations at the bank level, called
  \emph{per-bank refresh (\refpb)}, which is currently supported in LPDDR DRAM
  used in mobile platforms~\cite{jedec-lpddr3}. In contrast to \refab, \refpb
  enables a bank to be accessed while another bank is being refreshed,
  alleviating part of the negative performance impact of refresh.
  \figref{per-bank-refresh-timeline} shows pictorially how \refpb provides
  performance benefits over \refab from parallelization of refreshes and reads.
  \refpb reduces refresh interference on reads by issuing a refresh to Bank 0
  while Bank 1 is serving reads. Subsequently, it refreshes Bank 1 \new{while
    allowing} Bank 0 to serve a read. As a result, \refpb alleviates part of the
  performance loss due to refreshes by enabling parallelization of refreshes and
  accesses across banks.

Unfortunately, there are two shortcomings of per-bank refresh. First, refreshes
to different banks are scheduled in a strict round-robin order, as specified by
the LPDDR standard~\cite{jedec-lpddr3}. Using this static policy may force a
busy bank to be refreshed, delaying the memory requests queued in that bank,
while other idle banks are available to be refreshed. Second, \new{a bank that
  is refreshing} \emph{cannot} concurrently serve memory requests. \new{Hence,
  requests to a refreshing bank get delayed due to a ``refresh--access bank
  conflict.''}

We show that the negative performance impact of DRAM refresh becomes exacerbated
as DRAM density increases in the future. \figref{perf_loss_all} shows the
average performance degradation of all-bank/per-bank refresh compared to ideal
baseline without any refresh.\footnote{Our detailed methodology is described in
  Section 5 of our full paper~\cite{chang-hpca2014}.} Although \refpb performs
slightly better than \refab, the performance loss \new{due to refresh} is still
significant, especially as the density grows (16.6\% loss at 32Gb). Therefore,
\textbf{the goal} this work is to provide practical mechanisms to overcome
\new{the aforementioned} two shortcomings to mitigate the performance overhead
of DRAM refresh.

\figputGHS{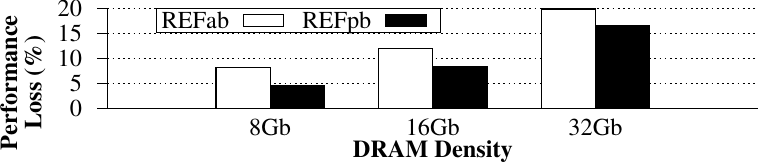}{1.0}{Performance loss due to \new{all-bank refresh}
  (\caprefab) and \new{per-bank refresh} (\caprefpb). Reproduced
  from~\cite{chang-hpca2014}.}

%
%
\section{Parallelizing Refreshes with \\ Memory Accesses}

We propose two mechanisms, \emph{\darplong (\darp)} and \emph{\intersub (\is)},
that hide refresh latency by parallelizing refreshes with memory accesses across
\emph{banks} and \emph{subarrays}, respectively. In this section, we present a
brief overview of these two new mechanisms. We refer the reader to Section 4 of
our \new{HPCA 2014} paper~\cite{chang-hpca2014} for more detail on the algorithm
and implementation.


\subsection{\darplong (DARP)}

\darp is a new refresh scheduling policy that consists of two components. The
first component is \emph{\ooolong}, which enables the memory controller to
specify a particular (idle) bank to be refreshed as opposed to the standard
per-bank refresh policy that refreshes banks in a strict round-robin order. With
out-of-order refresh scheduling, \darp can avoid refreshing (non-idle) banks
with pending memory requests, thereby avoiding the refresh latency for those
requests. The second component is \emph{\warplong} that proactively issues
\refpb to a bank while DRAM is draining write batches to other banks, thereby
overlapping refresh latency with write \new{request latencies}.

\subsubsection{DARP: Out-of-order Per-bank Refresh}

A major limitation of the current \refpb mechanism is that it disallows a memory
controller from specifying which bank to refresh. Instead, a DRAM chip has
internal logic that strictly refreshes banks in a \emph{sequential round-robin
  order}. Because DRAM lacks visibility into a memory controller's state (e.g.,
request queues' occupancy), simply using an in-order \refpb policy can
unnecessarily refresh a bank that has multiple pending requests to be served
when other banks may be \emph{free} to serve a refresh command. To address this
problem, we propose the first component of \darp, \emph{\ooolong}. The idea is
to remove the bank selection logic from DRAM and make it the memory controller's
responsibility to determine which bank to refresh. As a result, the memory
controller can refresh an idle bank to enhance parallelization of refreshes and
accesses, avoiding refreshing a bank that has pending requests as much as
possible.

Due to \refpb reordering, the memory controller needs to guarantee that
deviating from the original in-order \new{refresh} schedule still preserves data integrity. To
achieve this, we take advantage of the fact that the contemporary DDR JEDEC
standard~\cite{jedec-ddr4} provides some refresh scheduling flexibility. The
standard allows up to \emph{eight} all-bank refresh commands to be issued late
(postponed) or early (pulled-in). This implies that each bank can tolerate up to
eight \refpb commands to be postponed or pulled in. Therefore, the memory controller
ensures that reordering \refpb preserves data integrity by limiting the number
of postponed or pulled-in commands. Our \new{HPCA 2014} paper~\cite{chang-hpca2014} describes
our new algorithm for \ooolong in detail.

\subsubsection{DARP: Write-refresh Parallelization}

The key idea of the second component of \darp is to actively avoid refresh
interference on read requests and instead enable more parallelization of
refreshes with \emph{write requests}. We make two observations that lead to our
idea. First, {\em write batching} in DRAM~\cite{lee-tech2010} creates an
opportunity to overlap a refresh operation with a sequence of writes, without
interfering with reads. A modern memory controller typically buffers DRAM writes
and drains them to DRAM in a batch to amortize the \emph{bus turnaround
  latency}, also called \emph{tWTR} or
\emph{tRTW}~\cite{jedec-ddr4,kim-isca2012,lee-tech2010}, which is the additional
latency incurred from switching between serving writes to reads \new{and vice
  versa}. Typical systems start draining writes when the write buffer occupancy
exceeds a certain threshold until the buffer reaches a low watermark. This
draining time period is called the \emph{writeback mode}, during which no rank
within the draining channel can serve read
requests~\cite{chatterjee-hpca2012,lee-tech2010,stuecheli-isca2010}. Second,
DRAM writes are \new{usually} \emph{not} latency-critical because processors do
not stall to wait for them: DRAM writes are due to dirty cache line evictions
from the last-level cache\ch{~\cite{lee-tech2010,stuecheli-isca2010,seshadri-isca2014}}.

Given that writes are not latency-critical and are drained in a batch for some
time interval, they are more flexible to be scheduled with minimal performance
impact. We propose the second component of \darp, \emph{\warplong}, that
attempts to maximize parallelization of refreshes and writes. \Warplong selects
the bank with the minimum number of pending demand requests (both read and
write) and preempts the bank's writes with a per-bank refresh. As a result, the
bank's refresh operation is hidden by the writes in other banks.

\begin{figure}[h]
\centering
\subfloat[Scenario 1: Parallelize postponed refresh with writes.]{
    \begin{minipage}{\linewidth}
    \begin{center}
    \label{fig:inter-bank-postpone}
    \includegraphics[width=\linewidth]{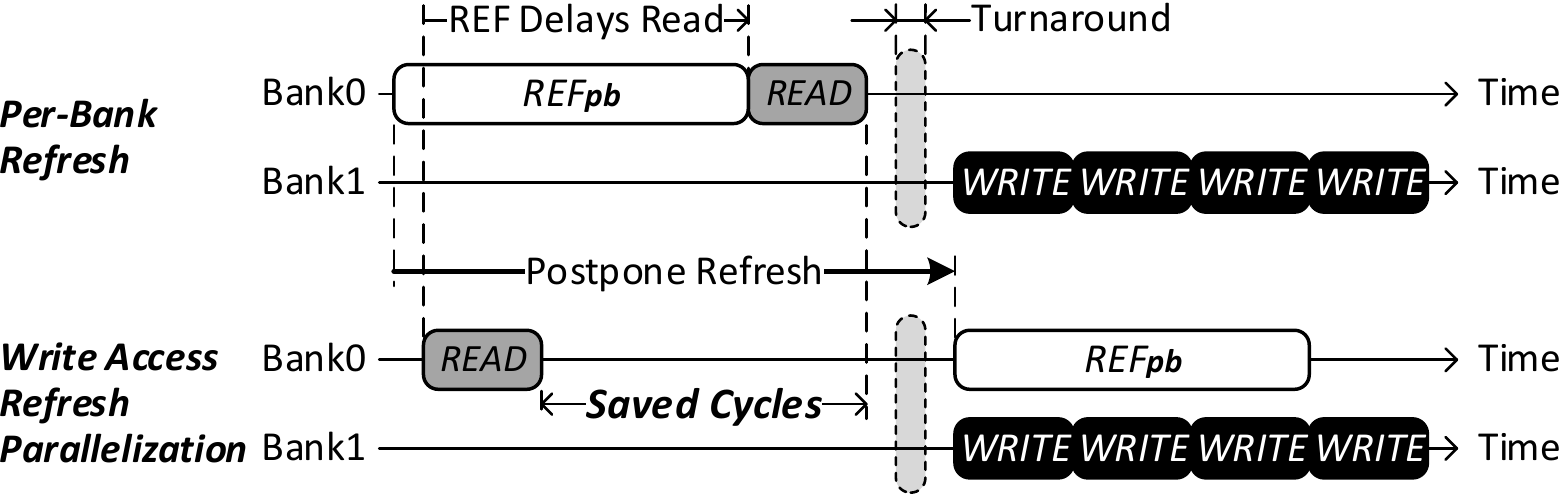}
    \end{center}
    \end{minipage}
}

\subfloat[Scenario 2: Parallelize pulled-in refresh with writes.]{
    \begin{minipage}{\linewidth}
    \begin{center}
    \label{fig:inter-bank-pull}
    \includegraphics[width=\linewidth]{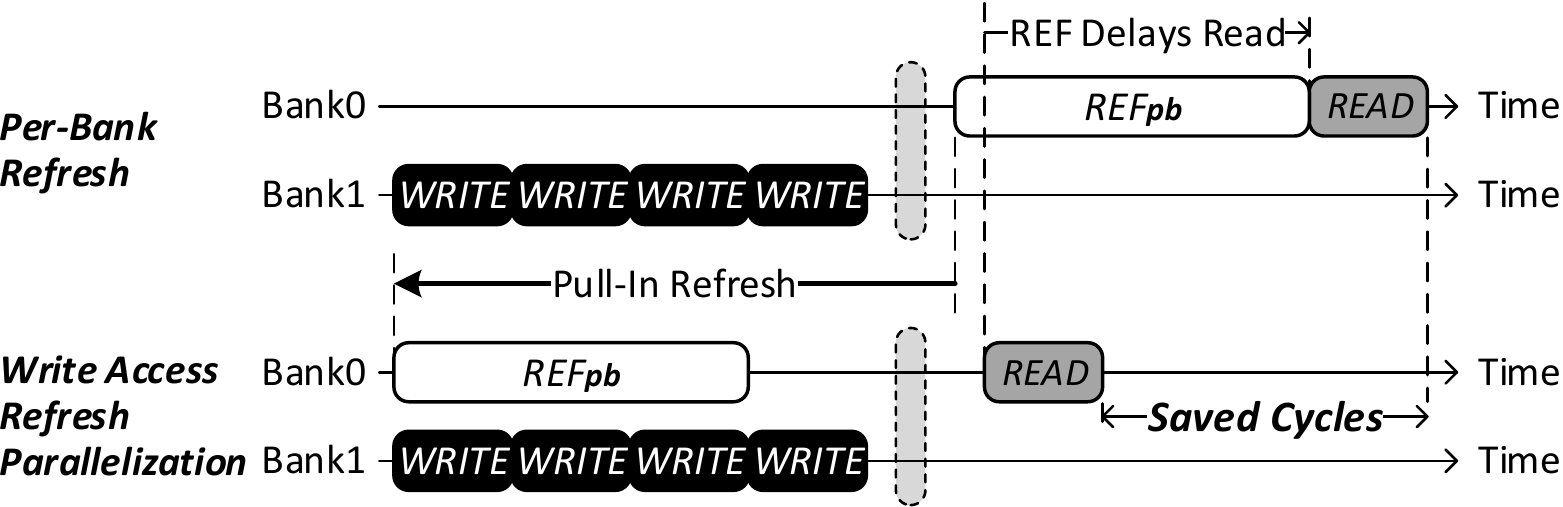}
    \end{center}
    \end{minipage}
}
\caption{Service timeline of a per-bank refresh operation along with read and
write requests using different refresh scheduling policies. Reproduced from
\cite{chang-hpca2014}.}
\label{fig:inter-bank-service-timeline}
\end{figure}

\figref{inter-bank-service-timeline} shows the service timeline and
benefits of \warplong. There are \textbf{two scenarios} when the
scheduling policy parallelizes refreshes with writes to increase
DRAM's availability to serve read
requests. \figref{inter-bank-postpone} shows the first scenario when
the scheduler \emph{postpones} issuing a \refpb command to avoid
delaying a read request in Bank 0 and instead serves the refresh in
parallel with writes from Bank 1, effectively hiding the refresh
latency in the writeback mode. Even though the refresh can potentially
delay individual write requests during writeback mode, the delay does
not impact performance as long as the length of writeback mode remains
the same as in the baseline due to longer prioritized write request
streams in other banks. In the second scenario shown in
\figref{inter-bank-pull}, the scheduler proactively \emph{pulls in} a
\refpb command early in Bank 0 to fully hide the refresh latency from
the later read request while Bank 1 is draining writes during the
writeback mode (note that the read request cannot be scheduled during
the writeback mode).

%
%
\subsection{\intersub (SARP)}

To tackle the problem of refreshes and accesses colliding within the same bank,
we propose \emph{\sarp (\intersub)}, which exploits the existence of
subarrays~\cite{kim-isca2012} within a bank. A DRAM bank is sub-divided into
multiple \emph{subarrays}\ch{~\cite{seshadri-micro2017, lee-hpca2013,
    lee-hpca2015, seshadri-cal2015, lee-sigmetrics2017, chang-hpca2016,
    hassan-hpca2016, kim-isca2012, seshadri-micro2013, vogelsang-micro2010,
    choi-isca2015, lu-micro2015,son-isca2013, yue-date2013, zhang-hpca2014}}, as
shown in \figref{bank-organization}. A subarray consists of a 2-D array of cells
organized in rows and columns.\footnote{Physically, DRAM has 32 to 128
  subarrays, which varies depending on the number of rows (typically 16-64K)
  within a bank. This work divides them into 8 \emph{subarray groups}. We refer
  to a subarray group as a subarray~\cite{kim-isca2012}, \new{without loss of
    generality}.} Each DRAM cell has two components: 1) a \emph{capacitor} that
stores one bit of data as electrical charge, and 2) an \emph{access transistor}
that connects the capacitor to a wire called \emph{bitline} that is shared by a
column of cells. The access transistor is controlled by a wire called
\emph{wordline} that is shared by a row of cells. When a wordline is raised to
$V_{DD}$, a row of cells becomes connected to the bitlines, allowing reading or
writing data to the connected row of cells. The component that reads \new{(i.e.,
  senses)} or writes a bit of data on a bitline is called a \emph{sense
  amplifier}, shared by an entire column of cells. A row of sense amplifiers is
also called a \emph{row buffer}. All subarrays' row buffers are connected to an
I/O buffer\new{~\cite{lee-pact2015, chatterjee-hpca2012, khoetal-isscc2009,
    moon-isscc2009}} that reads and writes data from/to the bank's I/O bus.

\figputHS{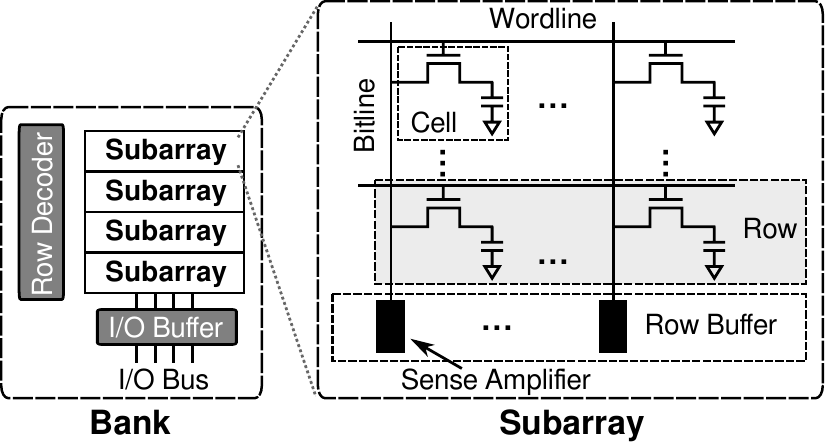}{0.8}{DRAM bank and subarray organization.
  Reproduced from~\cite{chang-hpca2014}.}

The key observation leading to our second mechanism, SARP, is that a refresh
operation is constrained to only a few \emph{subarrays} within a bank whereas
the other \emph{subarrays} and the \emph{I/O bus} remain idle during the process
of refreshing. The reasons for this are two-fold.  First, refreshing a row
requires only its subarray's sense amplifiers that restore the charge in the row
\emph{without} transferring any data through the I/O bus.  Second, each subarray
has its own set of \emph{sense amplifiers} that are \emph{not} shared with other
subarrays.

Based on this observation, SARP's key idea is to allow memory accesses to an
\emph{idle} subarray while other subarrays are refreshing.
\figref{sst} shows the service timeline and the
performance benefit of our mechanism. As shown, \is reduces the read latency by
performing the read operation to Subarray 1 in parallel with the refresh in
Subarray 0. Compared to \ib, \is provides the following advantages: 1) \is is
applicable to both all-bank and per-bank refresh, 2) \is enables memory accesses
to a refreshing bank, which cannot be achieved with \ib, and 3) \is also
utilizes bank-level parallelism\new{~\cite{lee-micro2009, mutlu-isca2008}} by
serving memory requests to multiple banks \new{in parallel} while the entire rank is
under refresh.

\figputHW{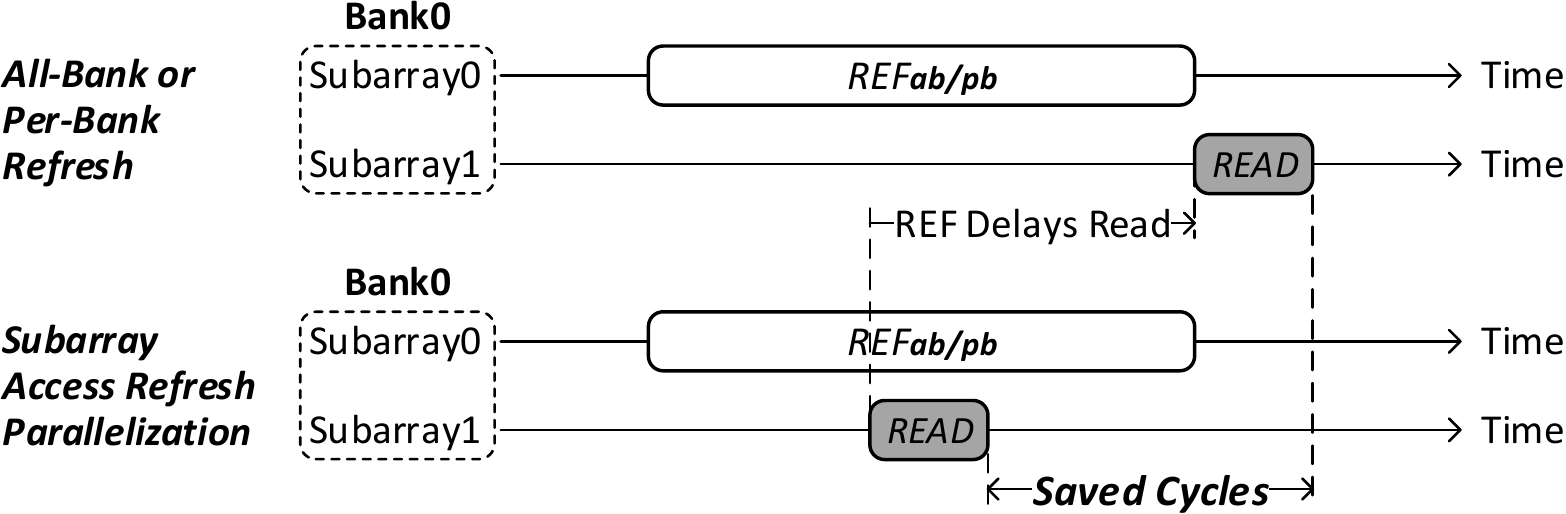}{Service timeline of a refresh and a read
request to two different subarrays within the same bank. Reproduced
from~\cite{chang-hpca2014}.}

\is requires modifications to 1) the DRAM architecture, because two distinct
wordlines in different subarrays need to be raised simultaneously \new{(to
  accommodate parallel refresh and access to \ch{the} two subarrays)}, which cannot
be done in today's DRAM due to the shared peripheral logic among subarrays; and
2) the memory controller, such that it can keep track of which subarray is under
refresh in order to send the appropriate memory request to an idle subarray.
Section 4.3 of our \new{HPCA 2014} paper~\cite{chang-hpca2014} describes these changes in
detail. To evaluate the benefits and die area overhead of SARP, we use 8
subarrays per bank and 8 banks per DRAM chip. Based on this configuration, we
calculate the area overhead of \is using parameters from a Rambus DRAM model at
55$nm$ technology~\cite{rambus_powermodel}, and find it to be 0.71\% in a 2Gb
DDR3 DRAM chip with a die area of 73.5$mm^2$. The power overhead of the
additional components is negligible compared to the entire DRAM chip.

\section{Evaluation}

We briefly summarize our results on an eight-core system. Section 6 of our
\new{HPCA 2014} paper provides detailed evaluations \new{on a wide variety of
  systems and sensitivity studies}. We evaluate the performance of our proposed
mechanisms on an eight-core system using Ramulator~\cite{ram-github,
  kim-cal2015}, an open-source cycle-level DRAM simulator, driven by CPU traces
generated from Pin~\cite{luk-pldi2005}. We use benchmarks from \emph{SPEC
  CPU2006~\cite{spec2006}, STREAM~\cite{stream}, TPC~\cite{tpc}}, and a
microbenchmark with random-access behavior similar to HPCC
RandomAccess~\cite{randombench}. Table~\ref{table:sys-config} summarizes the
configuration of our evaluated system.

\begin{table}[h]
\begin{footnotesize}
  \centering
    \setlength{\tabcolsep}{.55em}
    \begin{tabular}{ll}
        \toprule
\multirow{2}{*}{Processor} &
8 cores, 4GHz, 3-wide issue, 8 MSHRs/core,\\
& 128-entry instruction window
\\

        \cmidrule(rl){1-2}
\multirow{2}{*}{\begin{minipage}{0.5in}Last-level Cache\end{minipage}} &
64B cache-line, 16-way associative,\\
& 512KB private cache-slice per core \\

        \cmidrule(rl){1-2}
\multirow{3}{*}{\begin{minipage}{0.5in}Memory Controller\end{minipage}} &
64/64-entry read/write request queue, FR-FCFS~\cite{rixner-isca2000}, \\
& writes are scheduled in batches~\cite{chatterjee-hpca2012, lee-tech2010, stuecheli-isca2010}
with \\
& low watermark = 32, closed-row policy~\cite{chatterjee-hpca2012,
  kim-micro2010, rixner-isca2000, kim-hpca2010} \\

        \cmidrule(rl){1-2}
        \multirow{2}{*}{DRAM} & DDR3-1333~\cite{micronDDR3_8Gb}, 2 channels, 2 ranks per
        channel,\\
            & 8 banks/rank, 8 subarrays/bank, 64K rows/bank, 8KB rows \\

        \cmidrule(rl){1-2}

\multirow{2}{*}{\begin{minipage}{0.5in}Refresh Settings\end{minipage}}
        & $tRFC_{ab}$ = 350/530/890ns for 8/16/32Gb
        DRAM chips,\\
        & $tREFI_{ab}$ = 3.9\micro s, $tRFC_{ab}$-to-$tRFC_{pb}$ ratio = 2.3  \\

        \bottomrule
    \end{tabular}
  \caption{Evaluated system configuration. \new{Adapted from
      \cite{chang-hpca2014}.}}
  \label{table:sys-config}%
\end{footnotesize}
\end{table}

\figref{toppick-results} shows the average system performance (left) and energy
per DRAM access (right) of our final mechanism, \combo, the combination of \darp
and \sarp, compared to two baseline refresh schemes and an ideal scheme without
any refreshes. We measure system performance with the commonly-used
\emph{weighted speedup (WS)}~\cite{eyerman-ieeemicro2008,snavely-asplos2000}
metric. The percentage numbers on top of the bars are the performance
improvement of \combo over \refab.

\figputGHS{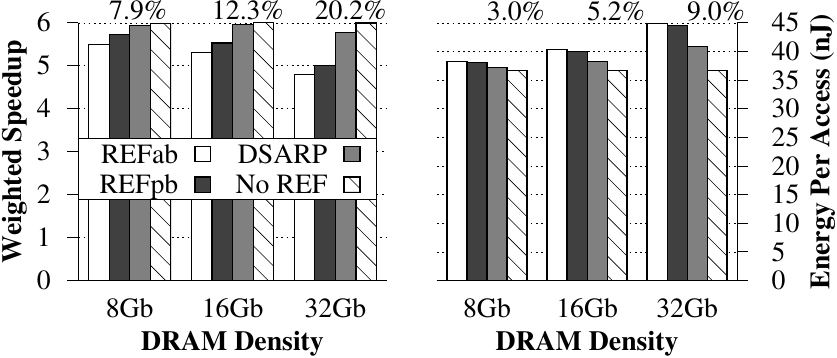}{1.0}{Average system performance and energy
  consumption \new{due to different refresh mechanisms}.}

We make two observations. First, \combo consistently improves system performance
and energy efficiency over prior refresh schemes, capturing most of the benefit
of the ideal \new{system with no refresh.} Second, as DRAM density (i.e.,
refresh latency) increases, the performance benefit of \combo gets larger.
\new{We conclude that DSARP is an effective mechanism to alleviate the negative
  performance impact of DRAM refresh.}

\subsection{Comparison to DDR4 \\ Fine Granularity Refresh}
\label{sec:ddr4}

DDR4 DRAM supports a new refresh mode called \emph{fine granularity refresh
  (FGR)} in an attempt to mitigate the increasing refresh latency
(\trfc)~\cite{jedec-ddr4}. \fgr trades off shorter \trfc with a faster refresh
rate ($\sfrac{1}{\trefi}$) that increases by either 2x or 4x. \figref{ddr4_ref}
shows the effect of \fgr in comparison to \refab, \emph{adaptive refresh policy
  (AR)}~\cite{mukundan-isca2013}, and \combo. 2x and 4x \fgr actually \emph{reduce}
average system performance by 3.9\%/4.0\%/4.3\% and 8.1\%/13.7\%/15.1\% compared
to \refab with 8/16/32Gb densities, respectively. As the refresh rate increases
by 2x/4x (higher refresh penalty), \trfc does \emph{not} scale down with the same
constant factors. Instead, \trfc reduces by 1.35x/1.63x with 2x/4x higher
rate~\cite{jedec-ddr4}, thus increasing the worst-case refresh latency by
1.48x/2.45x. This performance degradation due to \fgr has also been observed in
Mukundan et al.~\cite{mukundan-isca2013}.  AR~\cite{mukundan-isca2013}
dynamically switches between 1x (i.e., \refab) and 4x refresh modes to mitigate
the downsides of \fgr.  AR performs slightly worse than \refab (within 1\%) for
all densities. Because using 4x \fgr greatly degrades performance, AR can only
mitigate the large loss from the 4x mode and cannot improve performance over
\refab. On the other hand, \combo is a more effective mechanism to tolerate the
long refresh latency than both \fgr and AR as it overlaps refresh latency with
access latency without increasing the refresh rate.

\figputGHS{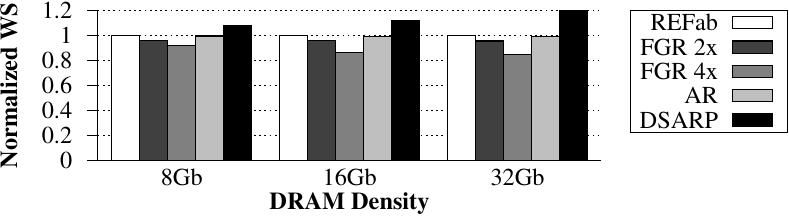}{1.0}{Performance comparisons to FGR and
  AR~\cite{mukundan-isca2013}. \new{Reproduced from \cite{chang-hpca2014}.}}

\new{We conclude that DSARP is an effective mechanism that can \ch{effectively tolerate and hide} longer
  refresh latencies, which are expected in future DRAM devices as DRAM technology
  scales to even smaller feature sizes.}

\section{Related Work}
\label{sec:related}

To our knowledge, this is the first work to comprehensively study the effect of
per-bank refresh and propose 1) a refresh scheduling policy built on top of
per-bank refresh and 2) a mechanism that achieves parallelization of refresh
operations and memory accesses {\em within} a refreshing bank. We discuss prior
works that mitigate the negative effects of DRAM refresh and compare them to our
mechanisms.

\textbf{Retention-Aware Refresh.} \new{Various} works (e.g.,~\cite{liu-isca2012,
  venkatesan-hpca2006,bhati-isca2015, lin-iccd2012, agrawal-memsys2016,
  nair-isca2013, gong-tc2016, patel-isca2017,
  kim-asic2001,baek-tc2014,agrawal-hpca2014,ohsawa-islped1998,qureshi-dsn2015})
propose mechanisms to reduce unnecessary refresh operations by taking advantage
of the fact that different DRAM cells have widely different retention
times~\cite{liu-isca2013, kim-edl2009, patel-isca2017}. These works assume that
the retention time of DRAM cells can be \emph{accurately} profiled and they
depend on having this accurate profile to guarantee data
integrity~\cite{liu-isca2013}. However, as shown in Liu et
al.~\cite{liu-isca2013} and later analyzed in detail by several other
works\new{~\cite{khan-sigmetrics2014, khan-dsn2016, khan-cal2016,
    patel-isca2017, qureshi-dsn2015, khan-micro2017}}, accurately determining
the retention time profile of DRAM is an outstanding research problem due to the
Variable Retention Time (VRT) and Data Pattern Dependence (DPD) phenomena, which
can cause the retention time of a cell to fluctuate over time. As such,
retention-aware refresh techniques need to overcome the profiling challenges to
be viable. A recent work, AVATAR~\cite{qureshi-dsn2015}, proposes a
retention-aware refresh mechanism that addresses VRT by using ECC chips, which
introduces extra cost. In contrast, our refresh mitigation techniques enable
parallelization of refreshes and accesses \emph{without} relying on cell data
retention profiles or ECC, thus providing high reliability at low cost.

\textbf{Refresh Scheduling.} Stuecheli et al.~\cite{stuecheli-micro2010} propose
elastic refresh that postpones refreshes by a time delay that varies based on
the number of postponed refreshes and the predicted rank idle time to avoid
interfering with demand requests. Elastic refresh has two shortcomings. First,
it becomes less effective when the average rank idle period is shorter than
\trfc as the refresh latency cannot be fully hidden in that period. This occurs
especially with 1) more memory-intensive workloads that inherently have less
idleness and 2) higher density DRAM chips that have higher \trfc. Second,
elastic refresh incurs more refresh latency when it {\em incorrectly} predicts a
time period as idle when \new{the time period} actually has pending requests. In
contrast, our mechanisms parallelize refresh operations with accesses even if
there is no idle period and therefore outperform elastic refresh.

Ishii et al.~\cite{ishii-msc2012} propose a write scheduling policy
that prioritizes write draining over read requests in a rank while
another rank is refreshing (even if the write queue has not reached
the threshold to trigger write mode). This technique is {\em only}
applicable in multi-ranked memory systems.  Our mechanisms are {\em
  also} applicable to single-ranked memory systems by enabling
parallelization of refreshes and accesses at the bank and subarray
levels, and they can be combined with Ishii et
al.~\cite{ishii-msc2012}.

Mukundan et al.~\cite{mukundan-isca2013} propose scheduling techniques (in
addition to adaptive refresh discussed in Section~\ref{sec:ddr4}) to address the
problem of {\em command queue seizure}, whereby a command queue gets filled up
with commands to a refreshing rank, blocking commands to {\em another}
non-refreshing rank.  In our work, we use a different memory controller design
that does not have command queues, similarly to prior
work~\cite{herrero-tc2013}. Our controller generates a command for a scheduled
request {\em right before} the request is sent to DRAM instead of pre-generating
the commands and queuing them up. Thus, our baseline design does not suffer
from the problem of command queue seizure.

\textbf{Subarray-Level Parallelism (SALP).} Kim et al.~\cite{kim-isca2012}
propose SALP to reduce bank serialization latency by enabling {\em multiple
  accesses} to different subarrays within a bank to proceed in a pipelined
manner. In contrast to SALP, our mechanism (\is) parallelizes {\em refreshes and
  accesses} to different subarrays within the same bank. Therefore, \is exploits
the existence of subarrays for a different purpose and in a different way from
SALP. We reduce the sharing of the peripheral circuits for refreshes and
accesses, not for arbitrary accesses. As such, our implementation is not only
different, but also less intrusive than SALP: \is does not require new DRAM
commands and timing constraints. \new{We note that several other works exploit
  the existence of subarrays for various performance and energy improvement
  purposes~\cite{lee-hpca2013, lee-hpca2015, chang-hpca2016, lee-sigmetrics2017,
    seshadri-micro2017, seshadri-cal2015, seshadri-micro2013}.}
\ch{We refer the reader to the SALP paper in this very same issue for a detailed
treatment of SALP~\cite{salp.tar18}.}

\textbf{DRAM Refresh Architecture.}
Several other works propose different refresh architectures. Nair et
al.~\cite{nair-hpca2013} propose Refresh Pausing, which pauses a refresh
operation to serve pending memory requests when the refresh causes conflicts
with the requests. Although our work already significantly reduces conflicts
between refreshes and memory requests by enabling parallelization, it can be
combined with Refresh Pausing to address rare conflicts. Tavva et
al.~\cite{tavva-taco2014} propose EFGR, which exposes non-refreshing banks
during an all-bank refresh operation so that a few accesses can be scheduled to
those non-refresh banks during the refresh operation. However, such a mechanism
does not provide additional performance and energy benefits over per-bank
refresh, which we use to build our mechanism in this dissertation. Isen and
John~\cite{isen-micro2009} propose ESKIMO, which modifies the ISA to enable
memory allocation libraries to skip refreshes on memory regions that do not
affect programs' execution. ESKIMO is orthogonal to our mechanism, and its
modification has high system-level complexity by requiring system software
libraries to make refresh decisions. \new{Other techniques (e.g., 
\ch{heterogeneous-reliability memory}~\cite{luo-dsn2014} or 
Flikker~\cite{liu-asplos2011}) can eliminate
  or reduce refreshes in parts of memory. Our techniques are \ch{complementary} to such
  refresh elimination/reduction techniques.}

\textbf{eDRAM Concurrent Refresh.} Kirihata et al.~\cite{kirihata-jssc2005}
propose a mechanism to enable a bank to refresh independently while another bank
is being accessed in embedded DRAM (eDRAM). Our work differs
from~\cite{kirihata-jssc2005} in two major ways. First, unlike
SARP,~\cite{kirihata-jssc2005} parallelizes refreshes only across banks, not
{\em within} each bank.  Second, there are significant differences between DRAM
and eDRAM architectures, which make it non-trivial to
apply~\cite{kirihata-jssc2005}'s mechanism directly to DRAM. In particular,
eDRAMs have no standardized timing/power integrity constraints and access
protocol, making it simpler for each bank to independently manage its refresh
schedule. In contrast, refreshes in DRAM need to be managed by the memory
controller to ensure that parallelizing refreshes with accesses does not violate
other constraints. \new{Other works (e.g., \cite{emma-ieeemicro2008,
    agrawal-hpca2013}) exploit the fact that eDRAM is used as a cache to avoid
  refresh operations.}

\section{Significance}

In this section, we describe three trends in the current and future DRAM
subsystem that will likely make our proposed solutions more important and
\new{attractive} in the future, and examine the work's impact on future research.

\subsection{Long-Term Impact}

\textbf{Worsening Retention Time.} As the DRAM cell feature size continues
to scale, the cells' retention time will likely become shorter, exacerbating the
refresh penalty\new{~\cite{mutlu-imw2013, mutlu-date2017, kang14}}. When the
surface area of cells gets smaller with further scaling, the depth/height of the
cell needs to increase to maintain the same amount of capacitance that can be
stored in a cell. In other words, the \emph{aspect ratio} (the ratio of a cell's
depth to its diameter) needs to be increased to maintain the capacitance.
However, many works have shown that fabricating high aspect ratio cells is
becoming more difficult due to processing technology~\cite{hong-iedm2010,
  kang14, mandelman-jrd2002}. Therefore, the cells' capacitance (and, thus,
their retention time) may potentially decrease with further scaling, increasing
the refresh frequency. Using DSARP is a cost-effective way to alleviate the
increasing negative impact of refresh as our results show~\cite{chang-hpca2014}.
\new{Note that errors have started appearing in DRAM chips due to aggressive
  technology
  scaling~\cite{kim-isca2014,mutlu-date2017,meza-dsn2015,sridharan-asplos2015,sridharan-sc2012,schroeder-sigmetrics2009}.
  The RowHammer problem is a prime example of DRAM errors that have been
  slipping into the field~\cite{kim-isca2014,mutlu-date2017}, and one solution
  for it is to increase the refresh rate~\cite{kim-isca2014,mutlu-date2017}.
  Such solutions to technology scaling issues clearly exacerbate the refresh
  problem. Therefore, DSARP can alleviate the performance impact under these
  conditions.}


\textbf{New DRAM Standards with Flexible Per-Bank Refresh.} According to
newer DRAM standards, the industry is already in the process of implementing a
similar concept of enabling the memory controller to determine which bank to
refresh. In particular, the two standards are: 1)
HBM~\cite{jedec-hbm,lee-taco2016} (October 2013, after the submission of our
HPCA 2014 paper~\cite{chang-hpca2014}) and 2) LPDDR4~\cite{jedec-lpddr4} (August
2014). Both standards have incorporated a new refresh mode that allows per-bank
refresh commands to be issued in \emph{any} order by the memory controllers.
Neither standard specifies a preferred order which the memory controller needs
to follow for issuing refresh commands.

Our work has done extensive evaluations to show that our proposed per-bank
refresh scheduling policy, \emph{DARP}, outperforms a naive round-robin
policy by opportunistically refreshing idle banks. As a result, our policy can
be potentially adopted in the future processors that use HBM or LPDDR4 DRAM.

\textbf{Increasing Number of Subarrays.} As DRAM density keeps increasing,
more rows of cells are added within each DRAM bank. To avoid the disadvantage of
increasing sensing latency due to longer bitlines in
subarrays~\cite{lee-hpca2013,chang-sigmetrics2016}, more subarrays
will likely be added within a single bank instead of increasing the size of each
subarray. Our proposed refreshing scheme at the subarray level, \emph{SARP},
becomes more effective at mitigating refresh as the number of subarrays
increases because the probability of a refresh and a demand request colliding
at the subarray level decreases with more subarrays.

\subsection{Potential Research Impact}

\textbf{Impact on Recent Research Work.} To our knowledge, this is the
first work to comprehensively study and extend the concept of \emph{per-bank
  refresh} to DDRx DRAM chips. Several works~\cite{bhati-isca2015,
  tavva-taco2014, guan-isqed2015} use our per-bank refresh mechanism as a
baseline for comparison. Kotra et al.~\cite{kotra-asplos2017} propose a new
refresh mechanism to further enhance our per-bank refresh mechanism. Kong et
al.~\cite{kong-microprocess2017} extend our per-bank refresh idea to eDRAM.

\textbf{Future Research Directions.} This work will likely create new
research opportunities for studying refresh scheduling policies at different
dimensions (i.e., bank and subarray level) to mitigate worsening refresh
\new{overheads}. Among many potential opportunities, one potential way to further
reduce the refresh latency (i.e., {\small\emph{$tRFC_{ab/pb}$}}) is to trade off
higher refresh rate (i.e., {\small\emph{$tREFI$}}), which is currently supported
as \emph{fine granularity refresh} in DDR4 DRAM for all-bank refresh. In this
work, we assume a fixed refresh rate for per-bank refresh as it is specified in
the standard. Therefore, a new research question that our work raises is
\emph{how can one combine per-bank refresh with fine granularity refresh and
  design a new scheduling policy for that}? We think that \darp can inspire new
scheduling policies to improve the performance of existing DRAM designs.

\new{
\textbf{Applicability to Other Memory Technologies.}
Refresh is used in NAND flash memory to improve
lifetime~\cite{cai-iccd2012,cai-hpca2015,cai-itj2013,luo-msst2015},
\ch{and can be used as a general solution to several other NAND flash reliability 
problems that are characterized and discussed in various recent 
works~\cite{luo-hpca2018, cai-dsn2015, cai-hpca2017, cai-date2013,
cai-date2012, cai-ieee2017, cai-ieeearxiv2017, cai-bookchapter2017,
cai-sigmetrics2014, luo-jsac2016}}.
We believe the idea of DSARP and refresh scheduling can also be applied to
refresh mechanisms in flash memory, and this can be especially beneficial toward
the end of the lifetime of flash memory when the device is refreshed more
frequently~\cite{cai-iccd2012,cai-ieee2017, cai-ieeearxiv2017, cai-bookchapter2017}. We refer the reader to our recent
works to understand the mechanisms for refresh in modern flash
memories~\cite{cai-ieee2017, cai-ieeearxiv2017, cai-bookchapter2017}.}

\ch{We believe the principles of DSARP are also applicable to emerging memory
technologies~\cite{meza-weed2013}, e.g., phase-change memory (PCM)~\cite{lee-isca2009,
lee-ieeemicro2010, qureshi-isca2009, yoon-taco2014,lee-cacm2010,qureshi-micro2009,
yoon-iccd2012, wong-ieee2010},
STT-MRAM~\cite{naeimi-itj2013, ku-ispass2013,guo-isca2009,chang-hpca2013},
or RRAM/memristors~\cite{chua-ieeetoct71, strukov-nature2008, wong-ieee2012}.
For example, PCM suffers from \emph{resistance drift}~\cite{wong-ieee2010, 
pirovano-ted2004, ielmini-ted2007}, where the resistance used to represent the value
becomes higher over time (and eventually can introduce a bit error).
To mitigate resistance drift, PCM can use refresh-like operations to rewrite
the original data value, and as the density of PCM grows, more such operations
are required.
We leave a detailed exploration of how DSARP can be used for emerging
memory technologies to future works.}



\section{Conclusion}

We introduced two new complementary techniques, \ib (\darplong) and
\is (\intersub), to mitigate the DRAM refresh penalty by enhancing
\emph{refresh--access parallelization} at the bank and subarray levels,
respectively.  \darp 1) issues per-bank refreshes to idle banks in an
out-of-order manner instead of issuing refreshes in a strict
round-robin order, 2) proactively schedules per-bank refreshes during
intervals when a batch of writes are draining to DRAM.  \is enables a
bank to serve requests from idle subarrays in parallel with other
subarrays that are being refreshed. Our extensive evaluations on a
wide variety of systems and workloads show that these mechanisms
significantly improve system performance and outperform
state-of-the-art refresh policies, approaching the performance of
ideally eliminating all refreshes. We conclude that \ib and \is are
effective at hiding the refresh latency penalty in modern and
near-future DRAM systems, and that their benefits increase as DRAM
density increases.

\new{We believe these techniques are also applicable to other memory
  technologies, such as NAND flash memory and phase change memory. We hope our
  work inspires future research to develop even more effective refresh latency
  tolerance techniques.}

\section*{Acknowledgments}

We thank Saugata Ghose for his dedicated effort in the preparation
of this article.
We thank the anonymous reviewers and Jamie Liu for helpful feedback and the
members of the SAFARI research group for feedback and the stimulating
environment they provide. We acknowledge the support of IBM, Intel, and Samsung.
This research was supported in part by the Intel Science and Technology Center
on Cloud Computing, \new{the Semiconductor Research Corporation, and an NSF CAREER
Award (grant 0953246)}.

\balance
{
\bstctlcite{bstctl:etal, bstctl:nodash, bstctl:simpurl}
\bibliographystyle{IEEEtranS}
\bibliography{kevin_paper}
}

\end{document}